\documentclass[conference]{IEEEtran}
\IEEEoverridecommandlockouts
\usepackage{cite}
\usepackage{amsmath,amssymb,amsfonts}
\usepackage{amssymb}
\usepackage{algorithmic}
\usepackage{graphicx}
\usepackage{textcomp}
\usepackage{xcolor}
\usepackage{comment}
\usepackage{subcaption}

\usepackage{algorithm}
\usepackage{mathtools}
\usepackage{enumitem}

\def\BibTeX{{\rm B\kern-.05em{\sc i\kern-.025em b}\kern-.08em
    T\kern-.1667em\lower.7ex\hbox{E}\kern-.125emX}}
\begin{document}

\title{TradeChain: Decoupling Traceability and Identity in Blockchain enabled Supply Chains \\
}
\author{}

\author{
\IEEEauthorblockN{
    Sidra Malik\IEEEauthorrefmark{1},
    Naman Gupta\IEEEauthorrefmark{2}
    Volkan Dedeoglu\IEEEauthorrefmark{3},
    Salil S. Kanhere\IEEEauthorrefmark{1},
    and Raja Jurdak\IEEEauthorrefmark{4}
}
\IEEEauthorblockA{
    \IEEEauthorrefmark{1}UNSW, Sydney
    \IEEEauthorrefmark{2}IIT, Delhi
    \IEEEauthorrefmark{3}CSIRO Data61, Brisbane
    \IEEEauthorrefmark{4}QUT, Brisbane \\
    \small \{sidra.malik, salil.kanhere\}@unsw.edu.au,
     naman.gupta1003@gmail.com,
    volkan.dedeoglu@data61.csiro.au, r.jurdak@qut.edu.au
}}


\maketitle

\begin{abstract}
Blockchain technology can provide immutability, provenance and traceability in supply chains. To utilize Blockchain's full potential, it is important to link supply chain events to the relevant entities for traceability and accountability purposes. Authorized participation is realised through consortium of various organisations. Transactions are verified by peer nodes pertaining to the consortium. Hence, privacy preservation of trade sensitive information such as trade flows and locations of production, storage and retail sites cannot be ascertained. 
In this work, we propose a privacy-preservation framework, TradeChain, which decouples the trade events of participants using decentralised identities. TradeChain adopts the Self-Sovereign  Identity (SSI) principles and makes the following novel contributions: a) it incorporates two separate ledgers: a public permissioned blockchain for maintaining identities and the permissioned blockchain for recording trade flows, b) it uses Zero Knowledge Proofs (ZKPs) on traders' private credentials to prove multiple identities on trade ledger and c) allows data owners to define dynamic access rules for verifying traceability information from the trade ledger  using access tokens and Ciphertext Policy Attribute-Based Encryption (CP-ABE). A proof of concept implementation of TradeChain is presented on Hyperledger Indy and Fabric and an extensive evaluation of execution time, latency and throughput reveals minimal overheads.
\end{abstract}

\begin{IEEEkeywords}
Self Sovereign Identity, privacy, permissioned blockchain, supply chain, traceability
\end{IEEEkeywords}

%
\IEEEpeerreviewmaketitle

\section{Introduction}

Blockchain technology has set a new paradigm for traceability and provenance in supply chains. Most blockchain based supply chain solutions are designed to be permissioned\footnote{https://hbr.org/2020/05/building-a-transparent-supply-chain} \cite{pbc8884089,bC8493157,malik2018productchain,cui2019blockchain}. Permissioned/consortium blockchains provide features such as authorised participation, traceability and accountability which are essential for digital supply chain systems. Although these features have increased the adoption of blockchain technology in the supply chain domain, exposure of trade secrets to consortium members poses risks to traders' privacy. This is primarily due to the inherent feature of permissioned blockchain designs whereby an audit trail of supply chain events is associated with the identity of the authorised traders. 
Permissioned blockchains are usually accessible to administrative bodies, i.e., consortium members and the  transaction validating nodes, which typically belong to various organisations in the consortium network. Since access to the ledger is pre-defined, any consortium member/validator satisfying the accessibility conditions may draw information from the ledger. Thus, the identity of participants and access rules play an important role in linking the trade flows and other important information constituting to trade secrets such as site locations of raw materials/ processing units, materials used in product manufacturing, list of suppliers, etc. In addition, privacy regulations such as the European Union’s General Data Protection Regulation (“GDPR”)\footnote{GDPR Privacy, https://gdpr.eu/data-privacy/} could impact the adoption of blockchain technology for supply chains due to privacy concerns. Thus, greater attention is required in designing blockchain-enabled systems which can provide privacy yet use the credible features of blockchain technology for providing traceability, provenance and auditing. \par  

Participants' privacy in blockchain enabled supply chains can be enhanced in two ways; keeping the data private or the identities private. If the data is kept private, traceability or provenance information cannot be drawn. Thus, this paper focuses on ensuring privacy through keeping the identities private. One may argue that permission-less blockchains, Bitcoin and Ethereum for example, already allow participants to take part anonymously, using pseudonyms. 
Further, proposals such as Zerocash\cite{sasson2014zerocash}, Monero\cite{sun2017ringct}, Mimblewimble\cite{fuchsbauer2019aggregate}, etc. thwart certain attacks on privacy (linking attack, address clustering, network analysis) and enhance participants' anonymity. Reliance on complete anonymity can be exploited for many malicious and criminal activities and thus compromises authorised participation, traceability and audit-ability. In permissioned blockchains, some privacy preserving approaches have been proposed, e.g., stealth addresses\cite{maouchi2019decouples}, Hidden Markov Models\cite{mitani2020traceability}, anonymous identities\cite{hardjono2019verifiable}. However, most of these methods do not provide fine-grained access control for audits or queries on privacy preserved data. Secondly, data collation is performed by default query mechanisms in permissioned blockchains which allows validators to query participants' data anytime without their authorisation. Hence, effective blockchain-supported supply chains  must support mechanisms that only allow validators or information requesters to link trade related information of participants with their explicit consent while simultaneously not comprising their identity. \par 

\begin{table*}[t] 
\caption{Comparison of privacy preservation approaches in permissioned blockchains}
  \label{tab:comp}
\begin{center}
\begin{tabular}{llccccccc}
\hline

\textbf{Article Identifier} & 
\textbf{Use-Case} & 
\multicolumn{1}{c}{\textbf{Privacy Mechanism}}& 
\multicolumn{1}{c}{\textbf{Platform}}& 
\multicolumn{1}{l}{\textbf{\begin{tabular}[c]{@{}l@{}}Identity\\  protection\end{tabular}}} & \multicolumn{1}{l}{\textbf{\begin{tabular}[c]{@{}l@{}}Data \\ protection\end{tabular}}} & 
\multicolumn{1}{l}{\textbf{\begin{tabular}[c]{@{}l@{}}End-End\\ Traceability\end{tabular}}} & 
\multicolumn{1}{l}{\textbf{\begin{tabular}[c]{@{}l@{}}Dynamic\\ Queries\end{tabular}}} &
\multicolumn{1}{l}{\textbf{\begin{tabular}[c]{@{}l@{}}Privacy \\ Analysis\end{tabular}}} \\ \hline

 Maochi et al. \cite{maouchi2019decouples}      & \begin{tabular}[c]{@{}l@{}}Supply Chains\\\end{tabular} &  \begin{tabular}[c]{@{}c@{}}Ring Signatures,\\ Range Proofs,\\ Stealth Addresses\end{tabular} & 
 \begin{tabular}[c]{@{}c@{}}Python based \\Permissioned BC\end{tabular}
 & Y
 & Y
 & Y
 & N
 & Y\\
 
Lu et al. \cite{lu2019blockchain}      & \begin{tabular}[c]{@{}l@{}}IIoT\\\end{tabular} &  
Federated learning & 
 \begin{tabular}[c]{@{}c@{}}Distributed\\Databases\end{tabular}
 & Y
 & Y
 & N
 & Y
 & Y\\
 
 Mitani et al. \cite{mitani2020traceability}      & \begin{tabular}[c]{@{}l@{}}Supply Chains\\\end{tabular} &  \begin{tabular}[c]{@{}c@{}}HMM, ZKP\\ Homomorphic\\ Encryption\end{tabular} & 
 \begin{tabular}[c]{@{}c@{}}Permissionless, \\Permissioned BC\end{tabular}
 & Y
 & Y(partial)
 & Y(partial)
 & N
 & Y\\
 
Androulaki\cite{androulaki2020privacy}      & \begin{tabular}[c]{@{}l@{}}Enterprise BC\\\end{tabular} &  \begin{tabular}[c]{@{}c@{}}Threshold\\ signatures,\\ Schnorr proofs, \\ commitment schemes\end{tabular} & 
 \begin{tabular}[c]{@{}c@{}}Permissioned BC\end{tabular}
 & Y(partial)
 & Y(partial)
 & Y
 & N
 & Y\\

Lin et al. \cite{lin2020ppchain}      & \begin{tabular}[c]{@{}l@{}}Supply Chains\\Cryptocurrency\end{tabular} &  \begin{tabular}[c]{@{}c@{}}Group \\ Signatures,\\ Broadcast,\\ Encryption\end{tabular} & 
 \begin{tabular}[c]{@{}c@{}}Permissioned BC\end{tabular}
 & Y
 & Y
 & Y(partial)
 & N
 & Y\\

Hardjono et al. \cite{hardjono2019verifiable}      & \begin{tabular}[c]{@{}l@{}}Blockchains\\\end{tabular} &  \begin{tabular}[c]{@{}c@{}}EPID ZKP,\\Membership\\ private keys\end{tabular} & 
 \begin{tabular}[c]{@{}c@{}}Permissioned BC\end{tabular}
 & Y
 & Y(partial)
 & N
 & N
 & N\\
 
Malik et al.      & \begin{tabular}[c]{@{}l@{}}Supply Chains\\\end{tabular} &  \begin{tabular}[c]{@{}c@{}}ZKP,\\ CP-ABE,\\ access tokens\end{tabular} & 
 \begin{tabular}[c]{@{}c@{}}Permissioned \\(Hyperledger Indy \\ -Fabric)\end{tabular}
 & Y
 & Y(partial)
 & Y
 & Y
 & Y\\
 \hline

\end{tabular}
\end{center}

\end{table*}

\begin{figure*}[t]
  \centering
  \includegraphics[width=16cm]{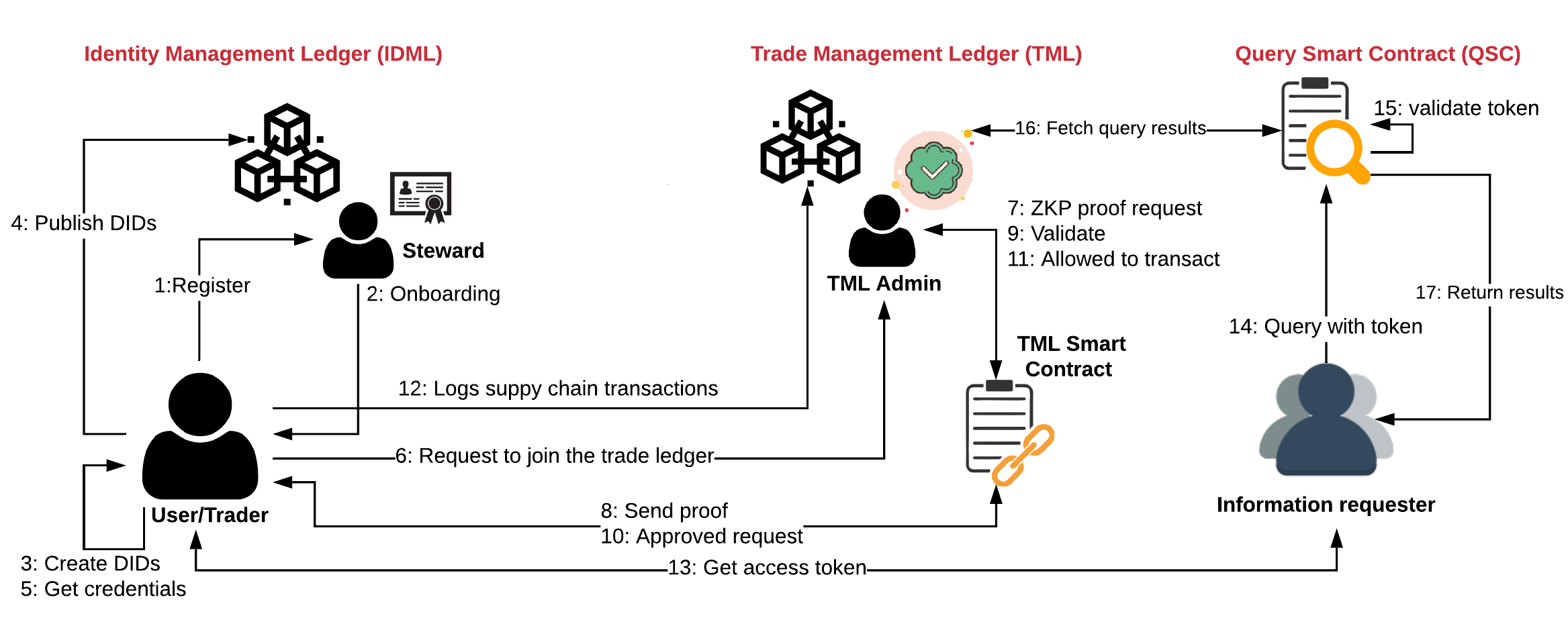}
  \caption{TradeChain Framework}
  \label{fig:overall}
\end{figure*}

In this work, we propose a privacy preserving framework called TradeChain, as described in Figure \ref{fig:overall}. TradeChain decouples the identity and trade events of supply chain traders by managing two separate ledgers: (i) Identity Management Ledger (IDML), a public permissioned blockchain for managing decentralised identifiers (DIDs) and (ii) Trade Management Ledger (TML), a permissioned blockchain for recording supply chain events. IDML leverages  privacy preserving and decentralised features of Self Sovereign Identities (SSIs)\footnote{https://sovrin.org/} \cite{tobin2016inevitable} where a trader can be given control to create and manage their identities using a digital wallet. For an authorised access to transact in TML \cite{malik2018productchain,malik2019trustchain}, the trader then uses the DIDs to prove his true identity by employing the concept of anonymous credentials ( based on Camenisch-Lysyanskaya signatures) and Zero Knowledge Proofs (ZKP) for credential verification \cite{camenisch2001efficient,camenisch2008efficient}. Our framework is designed to be compatible with privacy regulations such as GDPR so that the supply chain event history for a particular trader can only be collated from the TML using access tokens that are explicitly provided by that trader.
The main contributions of our work are: 
\begin{itemize}
    \item  a privacy preserving integrated framework of two separate ledgers IDML and TML for logging identities and trade activities, respectively. The framework allows supply chain entities to join TML using ZKPs on their credentials present on IDML and transact on TML using multiple decentralised identities; 
    \item a mechanism that allows supply chain entities on TML to define dynamic access rules for traceability of their data using access tokens and Ciphertext Policy based Encryption (CP-ABE); 
    \item a proof of concept implementation on Hyperledger Indy and Fabric where execution times, latency and throughput are benchmarked revealing minimal overheads.
\end{itemize}

\label{sec:intro}
It is worth mentioning that the TradeChain framework is applicable to any decentralised use-case apart from supply chains where privacy preservation, accountability, and user-centric fine grained access rules need to be realised simultaneously.

\section{Related Work}


In this section, we discuss some recent works in literature which aim to address the privacy preservation problem in blockchain enabled decentralised applications.\par

%
In \cite{maouchi2019decouples}, a traceability system called ``Decouples" is proposed to preserve privacy in blockchain based supply chain by incorporating different cryptographic techniques. Each actor in the system holds a certificate to transact on the ledger, while the owner of the certificate is hidden using a stealth address. For each new transaction, an actor creates a proof, indicating knowledge of his certificate using Elliptic Curve Integrated Encryption Scheme (ECIES). His identity is anonymized using Multi-Layered Linkable Spontaneous Anonymous Group (MLSAG) ring signatures. Moreover, the transaction amounts are protected using zero knowledge range proofs. To link the product specific information, the 'PASTA' protocol provides a single product-specific tracking key to track and reveal all the product-specific transactions of a particular actor. \par

In \cite{lu2019blockchain}, the authors design a permissioned blockchain-based secure data sharing architecture using federated learning. The data model is shared without actually revealing the data or the data owners.  When a query request is received, data owners use federated learning to train a data model. This model is based on the training set generated on local data query results. The information requester is then sent a model instead of data which is also locally cached by a permissioned blockchain node for future similar requests.\par

A similar problem of traceability and privacy protection is addressed in \cite{mitani8970301}. 
The authors aim to protect traceability related information such as: identity of trading participants, the amount of assets involved, the total amount transacted between permissioned blockchain and permission-less blockchain, and to check if some participants were involved or not in some of the transactions without revealing their identities. Hidden Markov Models (HMM), homomorphic encryption and zero knowledge proofs are used as privacy preservation mechanisms. The total number of additive operations while calculating the HMM,  correspond to the number of participants in the permissioned blockchain. The model is encrypted using homomorphic encryption and the model establishment is later verified by a protocol which uses non-interactive zero-knowledge proofs. \par

In\cite{androulaki2020privacy}, authors present a privacy-preserving token management system for permissioned blockchains that supports fine-grained auditing. A token is encoded using pedersen commitments where the token value, type and its owner is hidden. To prove that a user and token are registered on the ledger, ZK signature-based membership proofs are used. The validity of tokens is checked through trusted certifiers.\par

ChainAnchor\cite{hardjono2019verifiable} is an architecture for verifying anonymous identities in permissioned blockchains where entities can optionally disclose their identity pertaining to a specific transaction at the time of audit. First, Enhanced Privacy ID (EPID) zero-knowledge proof scheme is used for keeping the identities anonymous. Next, the entities are allowed to register their self-asserted transaction public-key to generate transactions in the permissioned ledger. Finally, the consensus nodes collectively enforce access control by only allowing transactions of the consortium members to be validated for participation.\par


PPChain \cite{lin2020ppchain} is another privacy preserving architecture which aims to provide anonymity and regulation in permissioned blockchains. PPChain is based on Ethereum and employs additional crypographic primitives such as group signatures and broadcast encryption. The consensus mechanism is replaced with the practical byzantine fault tolerance protocol which removes the transaction fee and mining reward.  Each transaction in PPChain is first cryptographically signed using Elliptic Curve Digital Signature Algorithm (ECDSA).  Since all the group members share the same public key, the specific group member signing the transaction cannot be identified, thus providing anonymity. The validating node verifies the transaction group signatures using the group public key. It then uses the private key of broadcast encryption to decrypt the ciphertext. The transaction is valid only if the ECDSA signature is correct, and the amount in the transaction is not more than the balance in the ledger.\par
The solutions discussed above address privacy challenges, however, they do not address the problem of dynamic access for linking the privacy preserved information. The read access for ledger is pre-defined and cannot be altered without updating the smart contract. Thus, the access rules cannot be dynamically changed with each query. Moreover, information access is not authorised by the data owner. The novelty of TradeChain stems from its ability to provide  identity protection and simultaneously allowing verification of traceability information only with the explicit consent of data owners. Table \ref{tab:comp} provides a comparison of TradeChain with the existing literature discussed in this section. TradeChain integrates two Hyperledger platforms, Indy and Fabric to maintain identity and trade information. The ability to support dynamic queries using access tokens from data owners, is a distinguishing feature of TradeChain.
\begin{figure}[t]
  \centering
  \includegraphics[width=\linewidth]{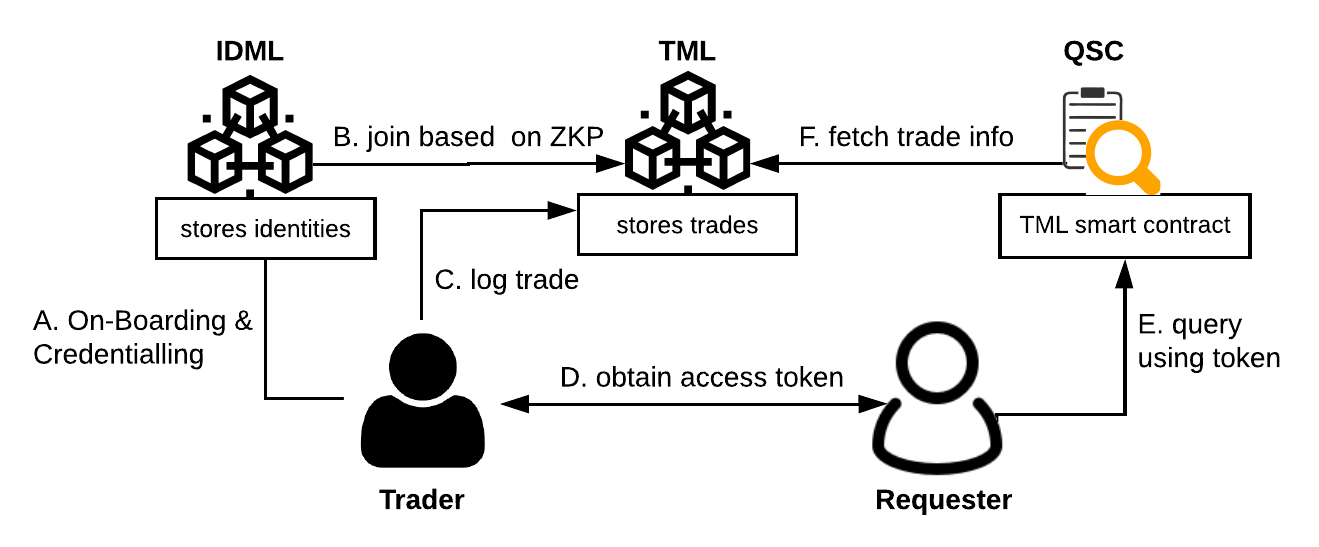}
  \caption{Overview of TradeChain}
  \label{fig:roadmap}
\end{figure}

\section{TradeChain Framework}
 In this section, we present an overview of the TradeChain framework where two distinct ledgers are used for decoupling the identity information from the trade history. For traceability purposes, we also discuss the query mechanisms based on access token issued by the traders, i.e. the data owners. A high-level overview of TradeChain is explained in Figure~\ref{fig:roadmap} using the phases A-F. There are three key components of Tradechain: Idenity Management Ledger (IDML), Trade Management Ledger (TML) and Query Smart Contract (QSC). IDML is a public permissioned blockchain based on Sovereign Identity Design\cite{tobin2018sovrin} which manages the data regarding the decentralised identities (DIDs) of SC entities. DIDs are publicly identifiable endpoints, such as documents, wallets, smart contracts or programmable agents\footnote{https://iop.global/what-is-ssi-did/}. 
 In phase A,  a supply chain trader first registers on IDML through an on-boarding process and obtains trader credentials. TML is a permissioned blockchain where only registered SC entities are allowed to participate. In phase B, the trader joins TML by proving his credentials acquired on IDML using Zero Knowledge Proofs (ZKPs). ZKP allows traders to prove credentials to TML admin without actually disclosing them. After the trader's credentials are proven, he is allowed to log trade related transactions on TML in phase C. For querying the trade transactions on TML, an information requester must first obtain access tokens from the relevant traders in phase D. The information requester next queries TML using these access tokens. QSC processes these queries by verifying the access tokens in phase E. In phase F, QSC fetches the relevant data from TML and returns the results to the requester without revealing traders' credentials or DIDs.\par

 In the following sections, each key component in Figure~\ref{fig:roadmap} is explained in detail. Section III-A gives a detailed overview of Tradechain. Section III-B discusses IDML, Section III-C presents TML and Section III-D gives the description of token based querying. Note, that the IDML design is based on key concepts of the open-sourced Sovrin project\cite{sovrin} (such as DIDs and their usage). However, in this work, the design is formulated in the context of supply chains and some of the underlying concepts are explained in detail to improve readers' understanding.

\subsection{Overview}
\label{sec:overview}

The architecture and interactions of the key components of Tradechain are shown in Figure~\ref{fig:overall}, where we expand on the phases A-F shown in Figure~\ref{fig:roadmap}. 
The trader first sends a registration request in step 1 (phase-A) to IDML followed by the execution of the on-boarding process in step 2 (phase-A). On-boarding involves registration of the trader on IDML using their private DIDs for communication between entities. Following this, the trader creates multiple DIDs in step 3 (phase-A) and depending on the type of DID, publishes them either on IDML or stores them in a wallet in step 4 (phase-A). The seller then obtains trader credentials from credential issuing authority using the DIDs in step 5 (phase-A). To trade in TML, the seller has to prove that he is a registered trader with IDML without having to disclose his full credentials. The seller first sends a join request to TML admin in step 6 (phase-B) who after accepting the request, responds with a zero knowledge proof request in step 7 (phase-B) which requires the seller to prove his trader credentials without disclosing them. The seller sends the proof in step 8 (phase-B). The TML admin validates the proof and approves the seller's registration request in TML in step 9-11 (phase-B). The seller can then log transactions on TML using different DIDs in step 12 (phase-C). Furthermore, to link the transactions in the TML to an individual seller, access tokens are issued by the traders (data owners) to the information requesters in step 13 (phase-D). The information requester issues a query transaction on TML using the access token in step 14 (phase-D). Based on the access rules specified in the token, a Query Smart Contract (QSC) validates the token in step 15 (phase-E) and fetches the trade information from TML in step 16 (phase-F). The query results are filtered by QSC to exclude any identity related information before they are sent to the information requesters in step 17 (phase-F).

\noindent \textbf{Credentials in Real World: }To further illustrate the use of DIDs in IDML, let us define how credentials are obtained and used in our daily life. In the real world, credentials represent the quality, achievement, skill, or qualification of an entity which indicates its suitability for a particular task. Credentials are defined through a hierarchy of administrative bodies such as credential issuers, certification authorities, government and regulatory bodies. For example, consider a driving licence as a  credential which is issued as a card and associated with a list of attributes such as name, age, address,etc. These list of attributes together form some structure, known as schema. A schema is a semantic structure which defines the standard for a list of attributes required for a credential. Schemas are defined by authorities allowing the credentials to be issued, for example, the Government defines a schema for a driving licence credential. The schema allows other subordinate organizations to issue credentials based on a standard. Thus, after the government has issued a schema for the driver's licence, a registration authority such as Transport NSW, Australia will verify Personally Identifiable Information (PII) of applicants, and conduct examinations to check the driving capability. It will then register them by assigning values to these schema attributes and issue the credentials in the form of driver's licence. The driver's licence will have some additional information apart from basic schema attributes such as name of issuing authority, validity etc. Similar to driver's licence credentials, every person owns several credentials to be used for different purposes. In the next section, we use these concepts of credentials, schemas and attributes in providing the digital credentials to entities on IDML.\par
\subsection{Identity Management Ledger (IDML)}
\label{sec:idml}
The purpose of IDML is to provide digital credentials to the traders using $DID$s. The DIDs on IDML can be verified without a need of any centralised authority\cite{indy_walk,sovrin}. IDML uses two types of DIDs: 1) A public DID known as "verinym", $DID_v$, must be publicly available and mapped to the real world identity of a supply chain entity 2) A private DID known as "pseudonym", $DID_p$ is used for online digital communication between two parties and kept private (in digital wallets). To maintain privacy, each SC entity can own multiple $DID_p$s for its trades. 
\par

\noindent \textbf{Digital Credentials and Identities in IDML:} 
\label{sec:dig-cred} The credentials on IDML are generated in a similar process as in real world using $DID_v$s and $DID_p$s of issuing authorities and SC entities. 
\begin{figure}[t]	\centerline{\includegraphics[width=\linewidth]{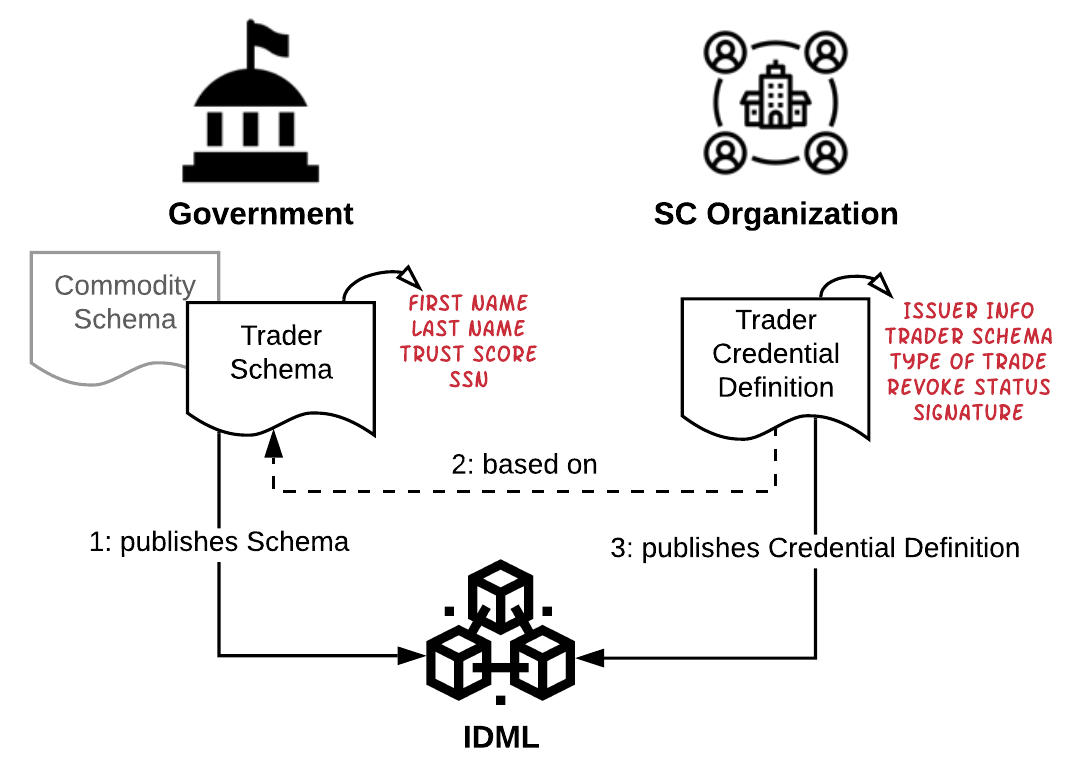}}
	\caption{Schema and Credentials}
	\label{fig:govsc}
	\vspace{-0.4cm}
\end{figure}
Consider an example of a simplified supply chain comprised of four entities: Government organization, Supply Chain Certification Authority (SCCA), a seller, and a buyer. The seller and buyer are interested in getting trader credentials from SCCA on IDML. These entities are publicly recognizable in IDML using their $DID_v$s and they communicate with each other to request credentials using $DID_p$s. Each $DID_v$ has an associated document which is generated from metadata fields of $DID_v$. The document contains the related information indicating service endpoints which can be used by any new entity to interact with an existing entity on IDML using $DID_p$. The roles and interactions of Government, SCCA, buyers and sellers are defined below and shown in Figure \ref{fig:govsc}:

\begin{itemize}
    \item Government is an organization responsible for recording schemas related to the credentials of supply chain entities in IDML, i.e. \textit{`trader'} schema, \textit{`commodity'} schema, etc. (see Figure \ref{fig:govsc} step 1).
    \item SCCA is responsible for registering the traders. The SCCA extends the trader schema by adding some additional attributes such as \textit{issuer info, type of trade, revocation status} (see Figure \ref{fig:govsc} step 2). These additional attributes together with the schema are called credential definition which is published on IDML (see Figure \ref{fig:govsc} step 3). SCCA verifies the traders based on their PII and trader schema, and issues trader credentials to the traders which entail the credential definition attributes and their respective values.
    \item Sellers and buyers are supply chain entities who interact with SCCA to get registered based on the trader schema.
    
\end{itemize}

Next, we explain the process of acquiring credentials, which entails three steps: (1)on-boarding, the process of getting $DID_p$, (2) publishing a verinym, the process of getting $DID_v$ and (3) credentialling, the process of acquiring credentials from SCCA.


\subsubsection{On-Boarding}
\label{sec:onboard}
  The process of establishing a connection between two entities on IDML using $DID_p$s is called on-boarding\cite{sovrin}. $DID_p$s are generated as a pair for every pairwise relationship. For example, if entity A and B have to communicate, entity A will generate a $DID_{p^{A-B}}$, and entity B will generate $DID_{p^{B-A}}$. $DID_p$s are shared and stored privately off-chain in the wallets rather than on IDML for confidentiality. Each $DID_p$ has an associated signing key and a verification key. The signing key is a private key and stored in the wallet along with the $DID_p$, whereas the verification key is a public key and stored on the IDML. Each entity must be on-boarded to IDML by an existing entity on IDML before they publish their $DID_v$s or request for credentials. Let the first default entity on IDML be referred to as the Steward. In real-life, Stewards can be a consortium of trusted governing bodies who would be responsible for running the IDML network and maintain credibility and trust in the system. These governing bodies could include entities such as Council of Supply Chain Management (CSCM), Supply Chain and Logistics Association of Australia (ACLAA), etc. However, Stewards have no central authority or control over DIDs of the supply chain entities. The process of on-boarding entities by a Steward is explained in Algorithm~1.
\begin{algorithm}[t]
\caption{On-Boarding Process}
\begin{algorithmic} [1]

\REQUIRE $DID_{v^{stwd}}$
\FOR{\textbf{each} entity $X \in IDML$} 
\STATE Initiate connection request with Steward
\STATE Steward generates $DID_{p^{S-X}}$ and saves it in his wallet
\STATE Steward logs $DID_{p^{S-X}}$ creation and $ver_{{key}^S}$ in IDML
\STATE Steward sends $con_{req} = [DID_{p^{S-X}} | ver_{{key}^{S-X}} | nonce)]$ to $X$
\STATE $X$ accepts $con_{req}$ and creates a wallet
\STATE $X$ creates $con_{resp} = [DID_{p^{X-S}} | ver_{{key}^{X-S}} | nonce]$ 
\STATE $X$ sends $Enc(con_{resp},ver_{{key}^{S-X}})$  to Steward 
\STATE Steward decrypts $con_{resp}$
\STATE Steward logs the creation of $DID_{p^{X-S}}$ and its $ver_{{key}^{X-S}}$ on IDML.
\ENDFOR
\end{algorithmic}
\end{algorithm}

\subsubsection{Publishing a Verinym}
 Once an SC entity is on-boarded, it can create a $DID_v$. An SC entity can create multiple $DID_v$s as required and can publish multiple schemas or credentials to IDML. 
An entity $X$ first creates a $DID_v$ in its wallet. It then prepares an encrypted message containing $DID_v$, and the corresponding verification key and sends it to Steward. Steward decrypts the message, queries the ledger for verification key of $DID_{p^{X-S}}$ and compares to the one in the decrypted message. If the verification keys match, Steward records $DID_v$ of $X$ on IDML.\par
Recall that, the Government is responsible for storing the trader schemas on the ledger. Thus, first the Government gets on-boarded by the Steward. After the Government acquires a $DID_v$ through the above mentioned process, it creates a trader schema and publishes it on the IDML. Anyone can verify that the schema is generated by the Government by verifying the $DID_v$ signatures in the associated document. In the next steps, we explain how this schema is used by the trader to obtain credentials.

\begin{figure*}[t!]
  \centering
  \includegraphics[width=15cm]{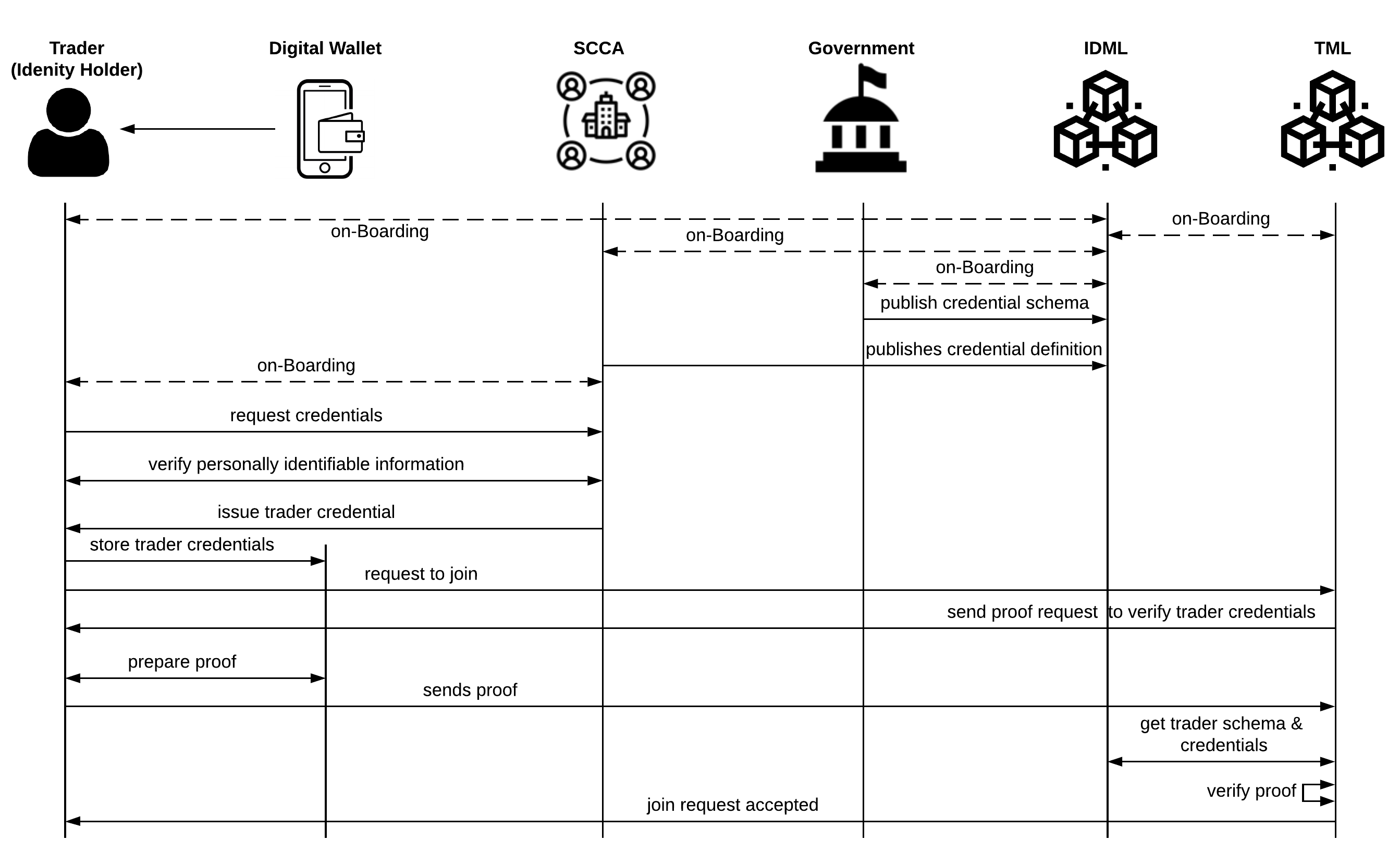}
  \caption{Credential Issuing}
  \label{fig:overall-credentials}
\end{figure*}

\subsubsection{Credentialling}
The \textit{credential definition} has two parts. The public part of the credential definition are the attributes which are published on IDML along with the $DID_v$ of the trader. The private part of the credential definition is the attribute of a trader which he wishes to conceal and is stored in his wallet. \textit{Credentials}, on the other hand, comprise of both attributes and their respective values .\par
Recall that SCCA publishes credential definitions on IDML, whereas the Government publishes the schemas. The trader after getting on-boarded on IDML, establishes a pairwise connection with SCCA using $DID_p$s and requests credentials from SCCA. SCCA issues the trader credentials after verifying the trader's eligibility. This process is outlined in Figure~\ref{fig:overall-credentials}.
It is important to note that the private credential definition in the wallet is not directly accessible except by the credential holder, i.e., the trader.
After getting the trader credentials, the trader can join TML using these credentials. 

\subsubsection{Transacting in Trade Management Ledger}
TML is a separate permissioned ledger governed by a consortium of supply chain entities such as earlier defined for Steward and proposed in \cite{malik2018productchain,cui2019blockchain}. The registration of new entities to TML is managed by the TML admin. Only those traders with verifable IDML credentials can record SC related transactions on TML. TML admin holds a $DID_v$ on IDML and processes the join requests of IDML traders to TML. In this process, TML admin requires a proof of trader credentials. TML allows the trader to choose any one of his $DID_v$s to get registered on TML. While registering, a seller needs to present only a proof of its credentials without disclosing the private credentials. \par
In the following steps, we explain in detail the process of a trader's registration in TML using DIDs, credentials and ZKP. The role of ZKP is to allow the trader to prove his trader credentials without fully revealing them. ZKP allows an entity (the Prover) to prove to another entity (the Verifier) that it has a knowledge of secret value, $X$, without disclosing any additional information. In our use case, the seller is the credential prover, while the TML admin is the credential verifier. The seller holds a credential, $C$, on his identity $X$ which asserts a certain property, $\mathcal{P}$ about $X$ which consists of attributes $m_1, m_2,\ldots, m_l$. The \textit{Prover} then presents $(\mathcal{P},C)$  to the \textit{Verifier}, which can verify that $C$ issued by the issuer (SCCA) has checked Prover's identity $X$ for a property $\mathcal{P}$. There are three main steps involved in the ZKP process: setup, proof generation and proof verification.
We give a brief description of each of these steps for enabling anonymous credentials\cite{indy_anoncred1,indy_anoncred2} based on the  Camenisch-Lysyanskaya signatures\cite{camenisch2001efficient}.

\noindent \textbf{Public Parameters:} 
All the entities in IDML are seeded with certain public parameters for executing the ZKP process. These parameters are generated as follows: 
\begin{enumerate}
\item generate a random 256-bit prime $\rho$ and a random 1376-bit number $b$ such that
$\Gamma = b\rho+1$ is prime and $\rho$ does not divide $b$;
\item generate random $g'< \Gamma$ such that
$g'^{b}\neq 1\pmod{\Gamma}$ and compute $g = g'^{b}\neq 1$.
\item generate random $r<\rho$ and compute $h = g^r$.
\end{enumerate}
Then $(\Gamma,\rho,g,h)$ are public parameters.

\noindent \textbf{Setup:} The ZKP setup must be executed for each Credential Issuer in the system.
Let $l$ be the number of attributes in $C$. Let $P$ be a description of the attribute set (types, number, length). Every credential is bound to a pseudonym $DID_p$ (between the credential issuer and holder), which is derived from the master secret, $m_1$ (s ee Hyperledger Indy\cite{indy_walk} ).
\begin{enumerate}
    \item generate random 1024-bit primes $p',q'$ such that  $p \leftarrow 2p'+1$ and $q \leftarrow 2q'+1$ are primes
    \item compute $n \leftarrow pq$
    \item generate a random quadratic residue $S$ modulo $n$
    \item select random $x_Z, x_{R_1},\ldots , x_{R_l}\in [2; p'q'-1]$ and
compute $Z \leftarrow S^{x_Z}\pmod{n} , R_i \leftarrow S^{x_{R_i}}\pmod{n}$ for $1\leq i \leq l$.
\end{enumerate}
The issuer's public key is $pk_I = (n, S,Z,R_1,R_2,\ldots,R_l,P)$ and the private key is $sk_I = (p, q)$. Next,  using the master secret $m_1$ and Issuer's public key $pk_1$, the prover generates a $DID_p$ to communicate with Issuer, $DID_{p^{P-I}}$ and stores credential $(\{m_i\},A,e,v)$ ( see \cite{indy_anoncred1} ) in his wallet using the process described earlier in Figure~\ref{fig:overall-credentials}. \par
\noindent \textbf{Proof Generation:} Let $\mathcal{A}$ be the set of \emph{all} attribute identifiers present in a credential. $\mathcal{A}_r$ are the identifiers of attributes that are revealed to the Verifier, and $\mathcal{A}_{\overline{r}}$ are those that are kept hidden by the Prover. A condition on hidden attributes such as ``an attribute $a > threshold$" can also be included within the proof request. The step by step process of proof generation is outlined in \cite{indy_anoncred1}.\par
The full proof $ \lambda$ is then sent to the Verifier which contains a sub-proof on credential $Pr_C$ and sub-proof for the condition $Pr_{con}$.

\noindent \textbf{Proof Verification:}
The Verifier uses Issuer's public key $pk_I$ involved in credential generation for the verification of the proof. Following Eq. 8-11 in \cite{indy_anoncred1}, the Verifier computes $\widehat{c}$. If $c=\widehat{c}$, the credentials are verified. Using the above mentioned ZKP steps, the trader is registered on TML using the following steps:
\begin{itemize}
    \item A trader in possession of trader credentials $C$, will initiate a connection with TML admin by executing the on-boarding process (See Section \ref{sec:onboard} and Algorithm 1). 
    \item After getting on-boarded, the trader sends a join request to TML admin.
    \item TML admin creates a proof request $\lambda_{req}$, which includes certain attributes and conditions to be met. For example, TML admin may request the seller to prove that: (i) the seller credentials $C$ are issued by the SCCA, (ii) his credentials are still valid (i.e., the revocation attribute of $C$ is false), and (iii) his reputation score, which is a private attribute in $C$, is greater than some threshold.
    \item The trader creates the proof $\lambda$ in response to the request $\lambda_{req}$ and sends it to TML admin.
   
    \item After receiving $\lambda$ from the seller, TML admin  retrieves credential schema, credential definition and $DID_v$  of the seller from IDML and verifies $\lambda$ according to the proof verification process described earlier. If $\lambda$ is verified, the trader gets registered on TML.
    \item The trader can create and trade new commodities using multiple $DID_v$s and $DID_p$s, respectively.
\end{itemize}

In the next sections, we explain in detail how trades are executed anonymously on the TML after the seller has registered on TML.
\subsection{Trade Management Ledger (TML)}
\label{sec:TML}
There are two main types of transactions stored on TML: (i) create transaction, $TX_{cr}$, for registering a commodity, and (ii) trade transaction, $TX_{tr}$, for trading a commodity between a seller and a buyer. The basic structure of these transactions is based on the framework presented in \cite{malik2018productchain}. 
The commodities are registered using $TX_{cr}$ where $DID_v$ serves as the identifier of the primary producer of the commodity. The $DID_v$ can be verified from the IDML. $TX_{tr}$ logs a trade of commodity between a seller and buyer using their pairwise $DID_p$s (see section~\ref{sec:onboard}). 
The structure of $TX_{cr}$ and $TX_{tr}$ is given below:
\begin{equation}
    TX_{cr}= [CID|H_{data}|DID_v|Sig_{S}]  
\end{equation}
\begin{multline}
    TX_{tr}= [CID|H_{data}|DID_{p^{S-B}}| \\DID_{p^{B-S}}|Sig_S|Sig_B]
\end{multline}
where $CID$ is the identifier of the commodity, $H_{data}$ is the hash of the commodity data (e.g. commodity type, quantity, unit price, etc.), $DID_v$ is the identifier of the seller. $Sig_{S}$ is the signature of the seller and $Sig_{B}$ is the signature of the buyer associated with the signing key of $DID_p$s. $DID_{p^{S-B}}$ is a pairwise DID from the seller to the buyer and $DID_{p^{B-S}}$ is the pairwise DID from the buyer to the seller. Both the DIDs are included to track the trades either at seller or buyer's end at the time of audit. Note that the owner of the commodity in this transaction is the seller, and the new owner is the buyer. A commodity will have only one $TX_{cr}$ but multiple $TX_{tr}$ through its supply chain journey. For anonymity, traders can create and use multiple $DID_v$s and  $DID_p$s  to log  commodities and trade on TML.


Before we explain our token based querying mechanism, it is important to summarise what is stored on IDML and TML:
\begin{itemize}
    \item \textbf{IDML} stores credential schemas, issuer public keys, $DID_v$s, verification keys, and revocation statuses, which are publicly accessible to anyone. However, $DID_p$s, credentials, proofs, and private keys are not published on the ledger and maintained off-chain in either wallets or other storage repositories. This private information cannot be accessed without the permission of information owner, i.e. trader.
    \item \textbf{TML} stores $TX_{cr}$s and $TX_{tr}$s, which are accessible to the traders logging these transactions. Other entities can retrieve the transaction history from the ledger only by using our token based querying mechanism. 
\end{itemize}

\subsection{Token based Querying}
Recall from Section \ref{sec:intro} and Section \ref{sec:overview}, using multiple $DID_p$s for logging $TX_{tr}$, which  makes it hard to link trade related transactions to individual traders. In this section, we describe in detail how to link the history of trade events to individual traders for the purpose of regulation and audits. We propose a token based query mechanism which allows traders to customise the access rules relevant to the query requests for their logged transactions.\par

 \begin{figure}[t]
  \centering
  \includegraphics[width=\linewidth]{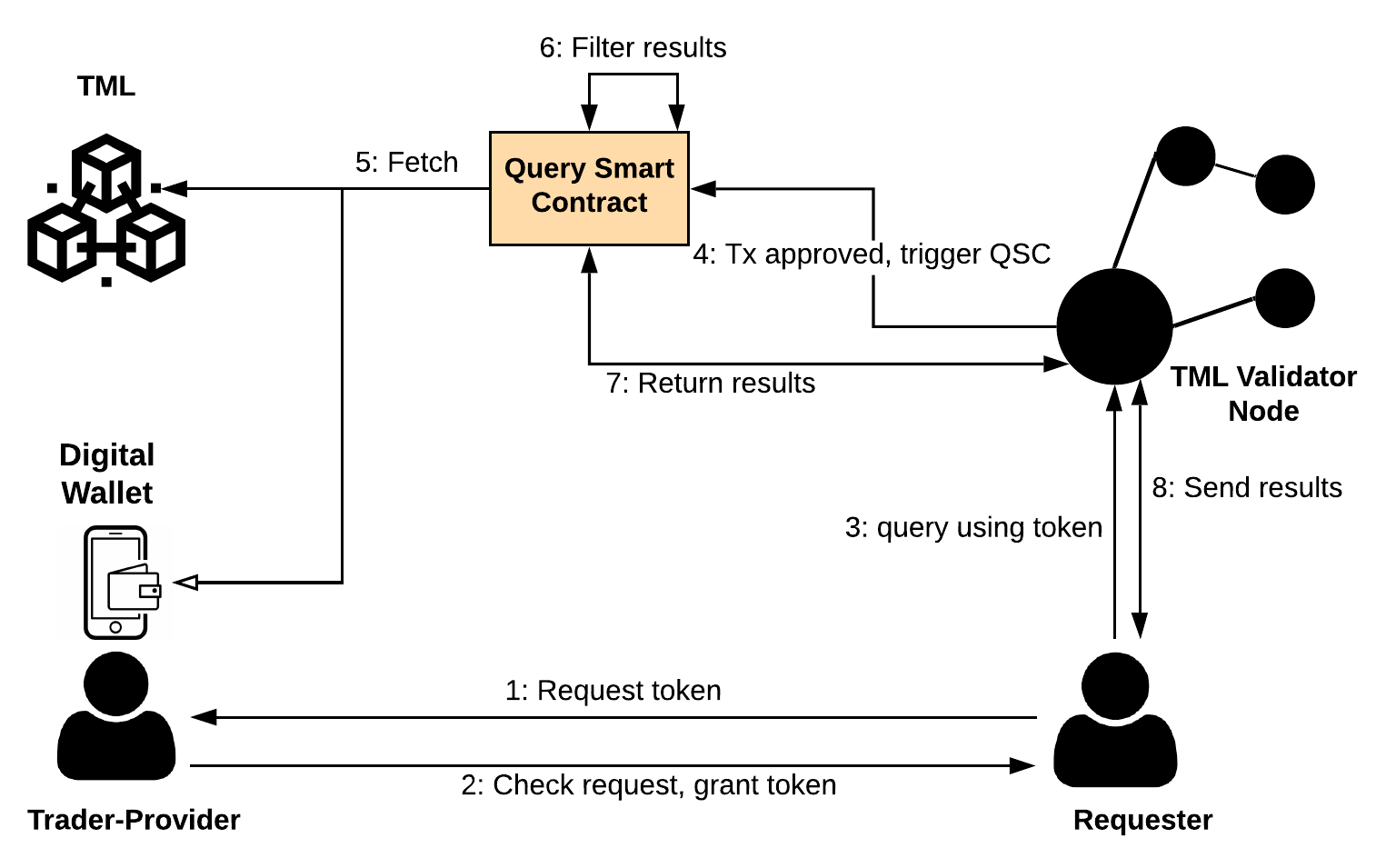}
  \caption{Token based querying process}
  \label{fig:qsc}
\end{figure}

The queries in permissioned blockhains are usually hard-coded in the chaincode, i.e., the query parameters to be returned in the final result are pre-defined. Once the query parameters are a part of a smart contract, they cannot be modified except when the smart contract is updated. For example, a query transaction which references a specific $TX_{cr}$, will result in all the parameters for that transaction type to be specified in the smart contract. Contrary to this traditional query mechanism, in TradeChain, we allow traders to reveal only partial information. For example, a trader may want to allow certain parameters $p_1$, $p_2$ to be accessible for an auditor of organization $A$ but $p_2$ not to be accessible for an auditor of organization $B$. In a permissioned ledger, granularity of data access is not dependent on the data owner but rather on the generic access rules defined in the smart contract. It would be cumbersome to add access rules specific to each trader and would require a smart contract to be updated every time a new access rule is added. Thus, our proposed token-based query mechanism for data access has the following goals:
\begin{itemize}
    \item Dynamic and granular access control: The mechanism should enable dynamic and granular access control of transactional data based on the consent of traders. 
    \item Anonymity: The mechanism should preserve the anonymity of traders by hiding DIDs associated with transactions while returning query results for transactional history.
\end{itemize}
Next, we describe how the above mentioned goals are met through token based access mechanism.

\subsubsection{Access Tokens}

The format of the access token $Token_{i,j}$ issued by a trader $i$ to a requester $j$ is defined as: 
\begin{multline}
    Token_{i,j} = [ID_{token}| Enc(R |Param_{i},{CP-ABE})|\\ Param_{j}|Time |validity |ID_j|H(Token_{i,j})_{Sig_i}]
\end{multline}
where $ID_{token}$ is the identifier of the token. $R$ specifies an end point for the smart contract to connect to trader's wallet API to allow access to $DID_p$s used for trade transactions. $Param_i$ and $Param_j$ are query parameters provided by the trader and requester respectively. These parameters must match for token validity. $Time$ specifies the time range for the queried data. The access token can be used multiple times in a specified time period. $validity$ specifies this time period and the number of times this token can be used for querying data. The token also includes the identity $ID_j$ of the requester $j$ (such as $DID_v$) and the  $H(Token_{i,j})$ is the hash of token signed with signature $Sig_i$ of the trader $i$. $Param_i$ and $R$ are encrypted using Ciphertext-Policy Attribute based encryption (CP-ABE)~\cite{waters-cpabe}. 
We use CP-ABE, mainly to encrypt the accessibility to $DID_p$s in $R$. CP-ABE allows a party encrypting the data to determine a policy for who can decrypt \cite{waters-cpabe}. The decryption policy is sent along with the ciphertext. This allows the decryption to be only dependent on policy rather than decryption keys or interaction with the party encrypting the data. Thus, traders can customise the access to their blockchain data using {CP\nobreakdash-ABE}.
\par
Access tokens will be acquired by the requester through an offline mechanism (e.g., peer-to-peer). It is important to note that although the trader's encryption policy is defined by the trader based on the parameters and the role of the requester (validator, consumer, trader, auditor), the policy must ensure that the decryption request is generated by QSC so that decryption is only possible within QSC and DIDs are not accessible outside QSC to validators of the blockchain system.

\smallskip
\subsubsection{Query Smart Contract (QSC)}
The QSC is triggered upon a query transaction which contains a valid access token. Figure \ref{fig:qsc} illustrates the steps for querying the permissioned ledger. 
In the first step, the requester sends a request for access token to the data provider, i.e. trader. The request specifies the query parameters such as number of trades and the time period for the record required. If the trader accepts the token request, he then sends the requester a valid token in step 2. Following that, the requester issues a query transaction with the access token to the blockchain validators in step 3. In step 4, the validators approve the transaction based on its format and the validity of the token. For example, a token is considered to be valid if the decrypted parameters match the $Param_j$ of the token and within the $validity$ and $Time$ period range (see Eq. 3). Once the transaction is validated, it triggers the QSC. In step 5, QSC will fetch the relevant information from the TML and trader's wallet API using $DID_v$ in the query transaction and $R$ from the decrypted token. In steps 6 and 7, QSC returns results by filtering the DIDs and parameters specified in the token. The filtered query results are then sent to the requester in step 8. In addition, the query transaction is also stored on TML specifying the requester and the hash of the token used for querying TML. The log of queries and associated tokens are kept to resolve any future conflicts related to data access.\par
\label{sec:arch}
\section{Evaluation and Results}
In this section, we first present the supply chain business model formulation of IDML and TML in the context of Hyperledger Indy and Fabric followed by the experimental setup. Next, we present results quantifying the performance of our system for relevant benchmarks. 
\subsection{Business Model}
We implement IDML and TML on Hyperledger Indy\footnote{https://www.hyperledger.org/use/hyperledger-indy} and Hyperledger Fabric\footnote{https://www.hyperledger.org/use/fabric}, respectively. For evaluation of IDML, we devise a commodity trading business network comprising: 
\begin{itemize}
\item Participants: traders (buyer or seller), Government, SC organization, regulator (see Section \ref{sec:idml}) . Government and SC organizations are part of IDML, traders are a part of IDML and TML, whereas regulators are information requesters who may not be a part of TML but have valid credentials in IDML. 
\item Assets: we define a simple commodity which could be any supply chain product either in its raw form or at its final production stage.
\item Smart Contracts: The trade smart contract functions handle the underlying functionality of $TX_{cr}$ and $TX_{tr}$. QSC consists of functions verifying access tokens and sending the filtered results.
\end{itemize}

\subsection{Experimental Setup}
\label{sec:expSetup}
The deployment of the business network and performance tests are carried out on a GPU Server (Intel(R) Xeon(R) CPU @ 3.70GHz, 6 cores, 65 GB memory). The business model setup for TradeChain involves the following setups: Hyperledger Indy for identity management, Hyperledger Fabric for trade management and Hyperledger Fabric Software Development Kit (SDK) for client application interaction with the chaincode.\par 
\subsubsection*{Network details}
We build a Fabric network of two organizations (seller and buyer) consisting of two peer nodes, one orderer (using SOLO as the ordering method), and a database (goleveldb). A seller and a buyer can start trading on TML after their DIDs have been verified from IDML and they are registered to trade in TML. Once authorised, the client application of these sellers and buyers can invoke registering or trading the commodities as per Eq. 1 and Eq. 2. \par

\subsubsection*{Packages and Libraries}
To integrate both IDML and TML, we setup two separate containers for Hyperledger Indy and Hyperledger Fabric.  Steward was created on Hyperledger Indy and the Admin was created on the Hyperledger Fabric network. Every time a new trader is created, two separate files are created in the trader's wallet: a) User Indy json folder- containing all $DID_v$s and $DID_p$s, b) Fabric X.509 certificate- issued by the certificate authority of Fabric after the trader credentials have been verified from IDML. Multiple $DID_v$s are created using the Indy \textit{nym} transaction and \textit{verinym} function which logs the $DID_v$ on IDML and stores it in the trader's wallet. After the $DID_v$ is used for a transaction, a new $DID_v$ is generated for the next transaction. The access tokens are created using json web tokens\footnote{https://www.npmjs.com/package/crypto-js}. The token is encrypted according to a CP-ABE policy using Charm-crypto library \footnote{http://charm-crypto.io/}. The token creation and validation are logged on the ledger by recording the information provider, requester, token hash, and a tag specifying either the token is ``generated" or ``submitted \& validated”. The proof-of-concept implementation and the functions for the above mentioned steps can be found online~\footnote{https://github.com/hyperledger/caliper-benchmarks}.



\begin{figure*}[t]
 \begin{center}
 \begin{tabular}{ c c }

\resizebox{0.40\textwidth}{!}
{\includegraphics{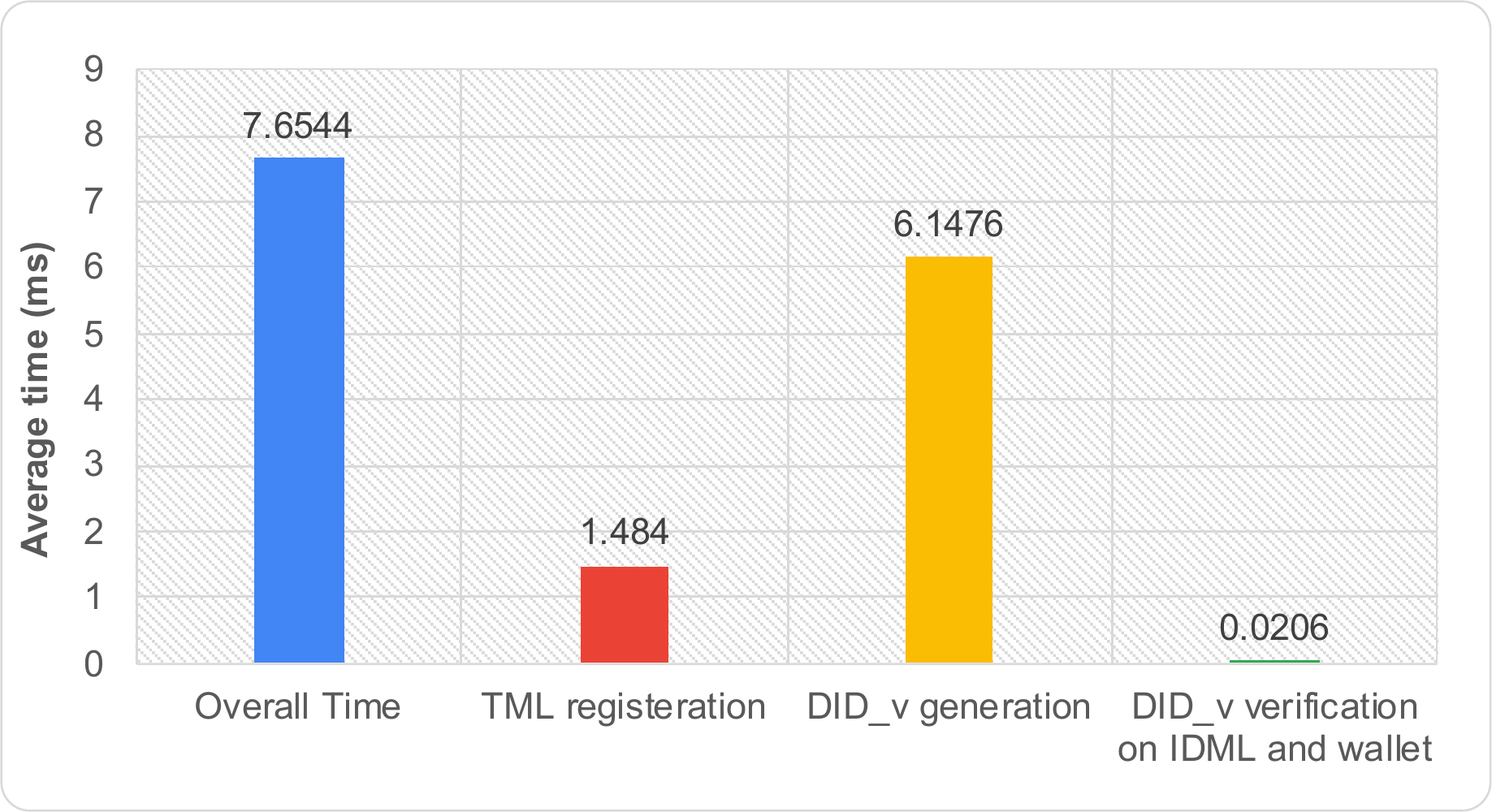}}     &
\resizebox{0.40\textwidth}{!}
{\includegraphics{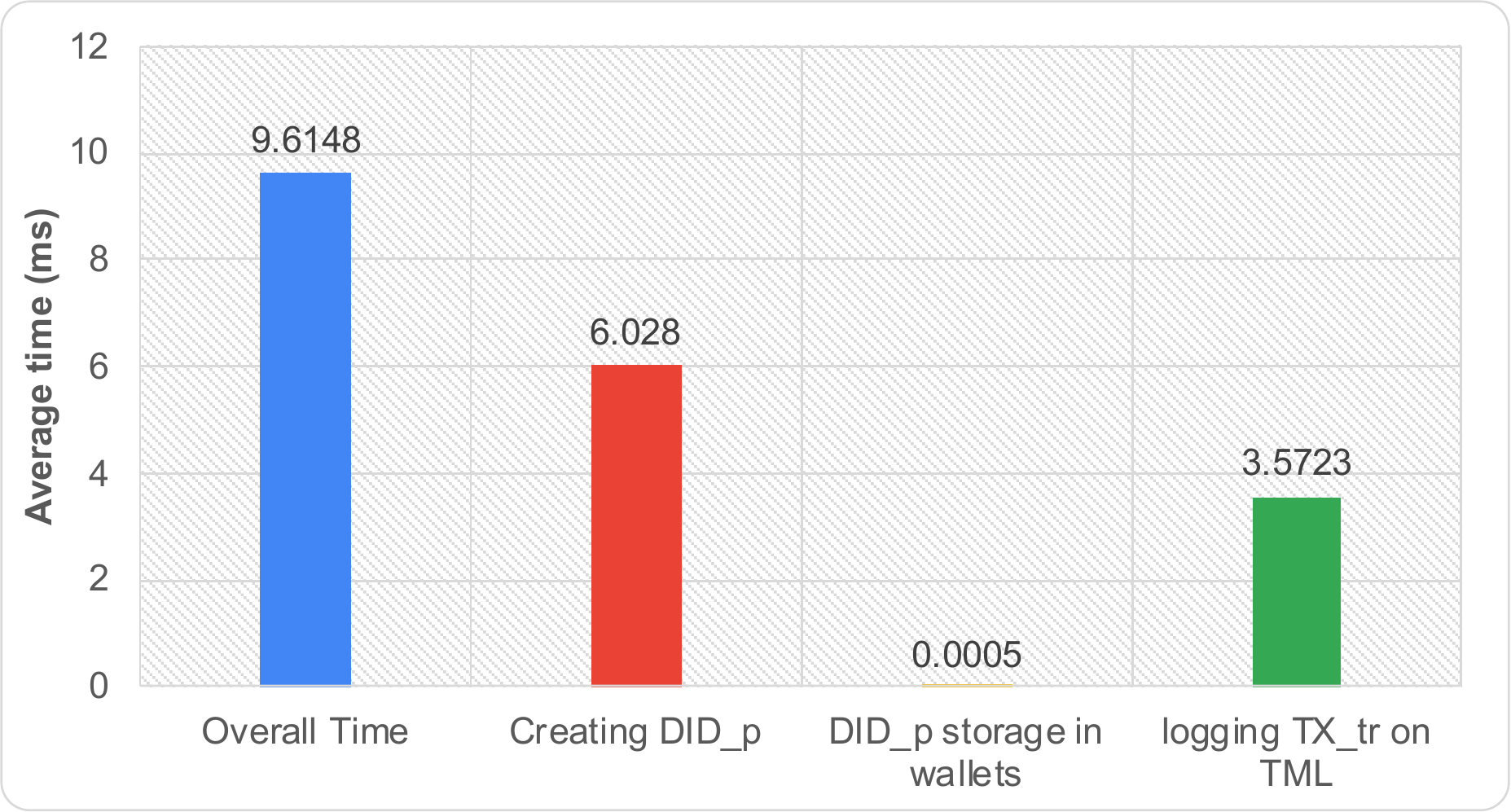}}  \\

{\small{(a) Trader's Registration on TML}} & 
{\small{(b) Trading a Commodity}} \\


\resizebox{0.40\textwidth}{!}
{\includegraphics{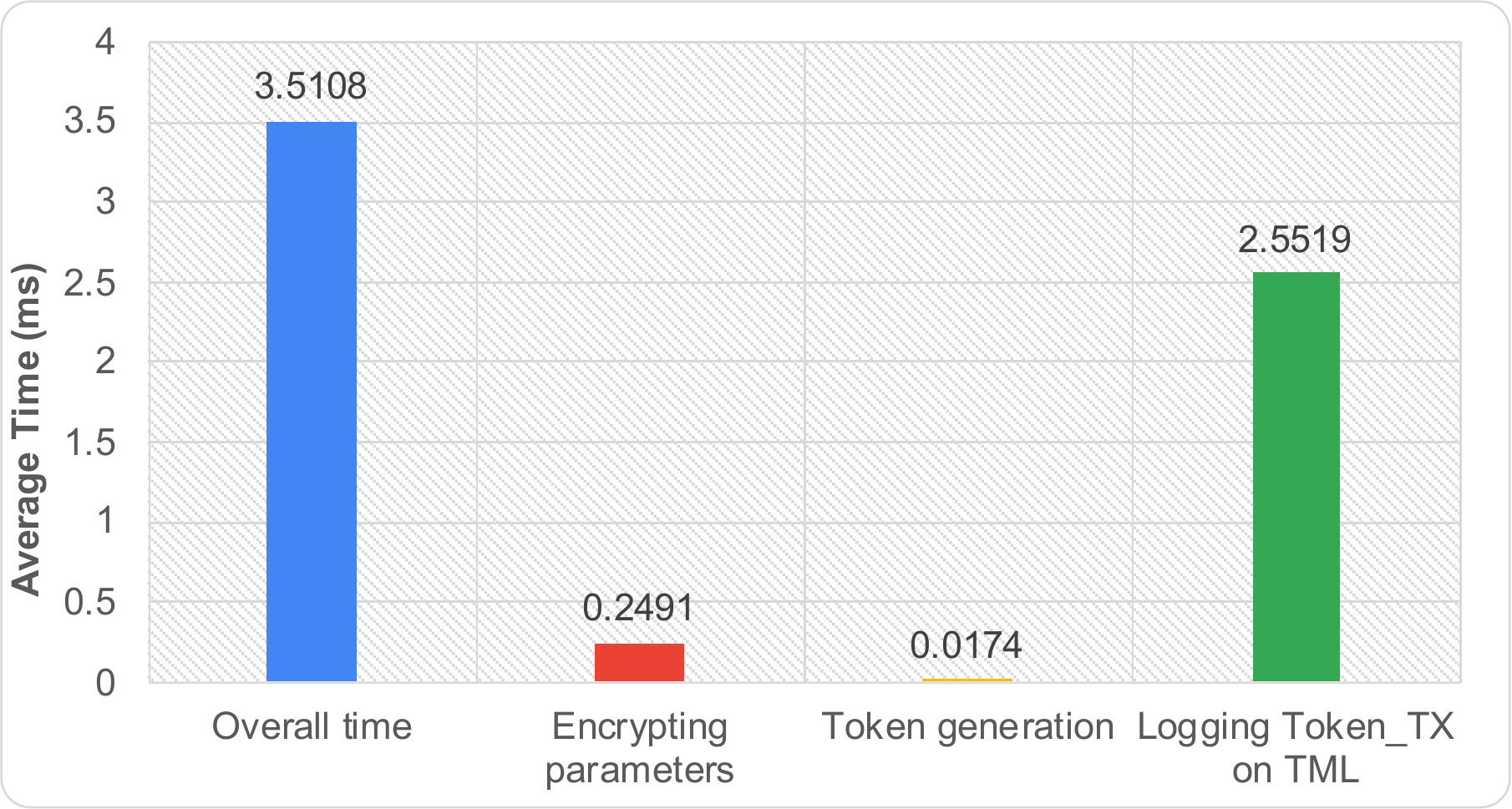}}  &
\resizebox{0.40\textwidth}{!}
{\includegraphics{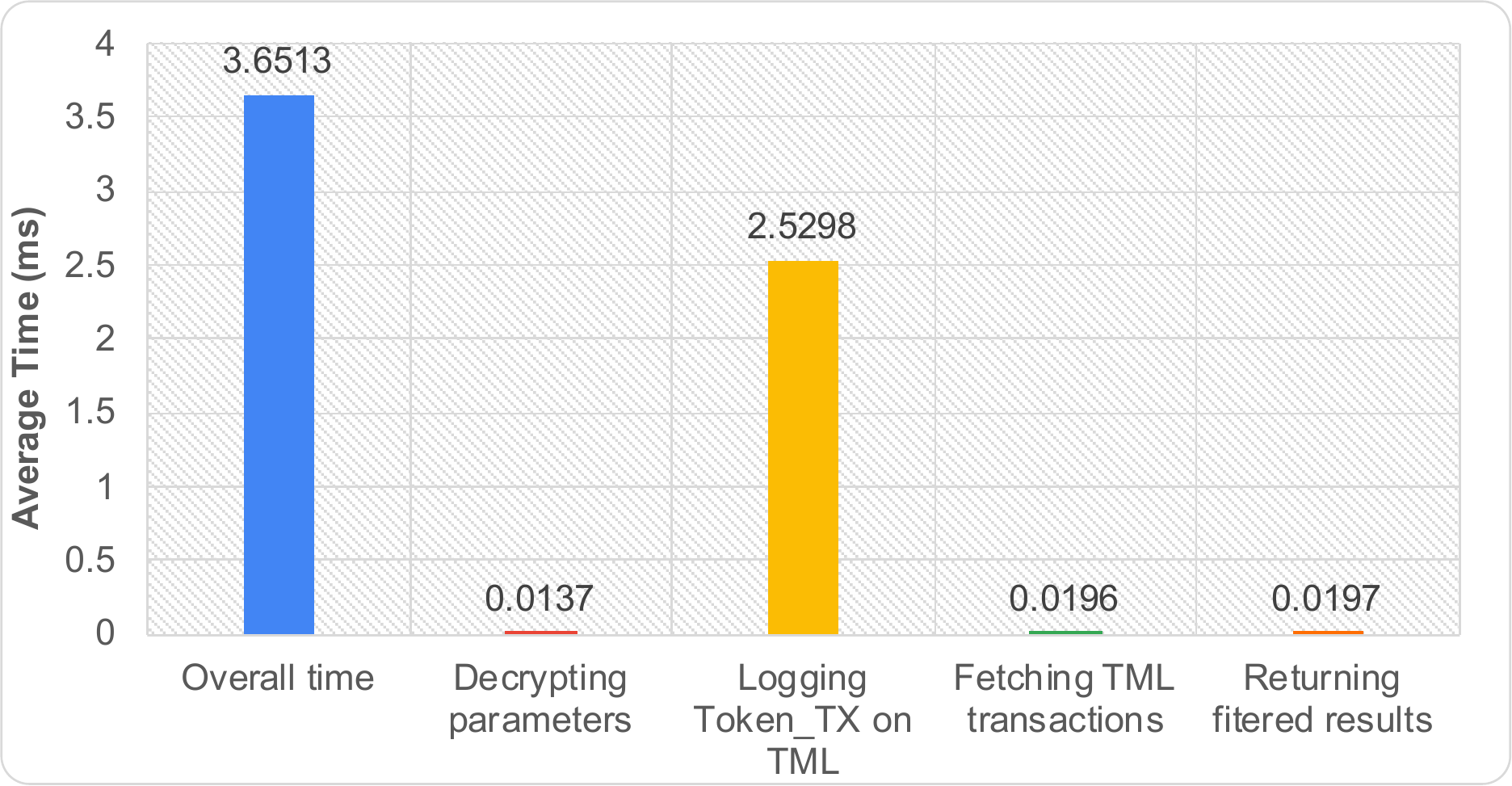}}   \\

{\small{(c) Generating a token}} &
{\small{(d) Validate token and return results}} \\ \small{{trading a commodity}}\\
\end{tabular}
\end{center}

\caption{Time Overheads of TradeChain}
\label{fig:timeres}
\end{figure*}
\subsection{Performance Evaluation}
We consider three metrics for TradeChain performance evaluation as described below:
\begin{itemize}
    \item Time overhead: refers to the processing time for transactions and functions involved in the IDML and TML. This time is measured from when a specific request is received at the smart contract until the appropriate response is sent back to the client.
    \item Latency: is the time taken from an application sending the transaction to the time it is committed to the ledger.
    \item Throughput: refers to the rate at which transactions are
committed to the ledger after they have been issued.
\end{itemize}

In the following text, we first measure the time overhead for all the steps defined in  Section \ref{sec:expSetup}. We then use Caliper to analyse the throughput and latency by increasing the transaction send rate on TML. All the results are  averaged across 10 runs in carrying out these computations.\par

\subsubsection{Time Overheads}
Figure \ref{fig:timeres} presents the time overheads of TradeChain functions. These functions include: $DID_v$ generation (by the trader), trader registration (by the TML admin), $DID_p$ generation and trading a commodity (by the trader), generating an access token (by the trader), and validating the access token and returning results (by the QSC).\par

Figure \ref{fig:timeres}-a shows trader's  registration on TML. The registration process involves $DID_v$ creation, its verification and credential proof verification by TML admin. The overall computation time for trader's registration is 7.65ms where $DID_v$ creation is the most time consuming step with an average time of 6.148ms. 
Trading a commodity involves generating a new $DID_p$, storing it in the trader's wallet and logging $TX_{tr}$ using the pair of $DID_p$s. The overall time for all these steps is 9.61ms as shown in Figure \ref{fig:timeres}-b. Generating a pair of $DID_p$s takes the most time, 6.028ms. In Figure \ref{fig:timeres}-c and Figure \ref{fig:timeres}-d we depict the time taken to generate the access token and return filtered results. It is observed that the time for validating token and returning results is 3.65ms which is quite low as compared to logging a single $TX_{cr}$ or $TX_{tr}$ on TML. This is because query transactions involve reading from TML ledger. In contrast, logging a $TX_{cr}$ and $TX_{tr}$ involves writing to the ledger with additional time for transaction endorsement, etc. In addition, if the number of query parameters are more, the time for validating the token and returning results is likely to increase. 

\subsubsection{Throughput and Latency Analysis}
In this section, we present the throughput and latency performance
evaluations on TML using Hyperledger Caliper. Hyperledger Caliper is a benchmark tool for latency and throughput analysis with its suitability to Fabric. Since Caliper does not support Hyperledger Indy, for IDML, a python test script can be used to send transactions manually to the network using the Indy API and measure the throughput and latency. However, we may expect slower results in this case due to the additional time associated with generating transactions from the script and receiving results within it.\par 

\begin{figure*}[h]
 \begin{center}
 \begin{tabular}{ c c }

\resizebox{0.42\textwidth}{!}
{\includegraphics{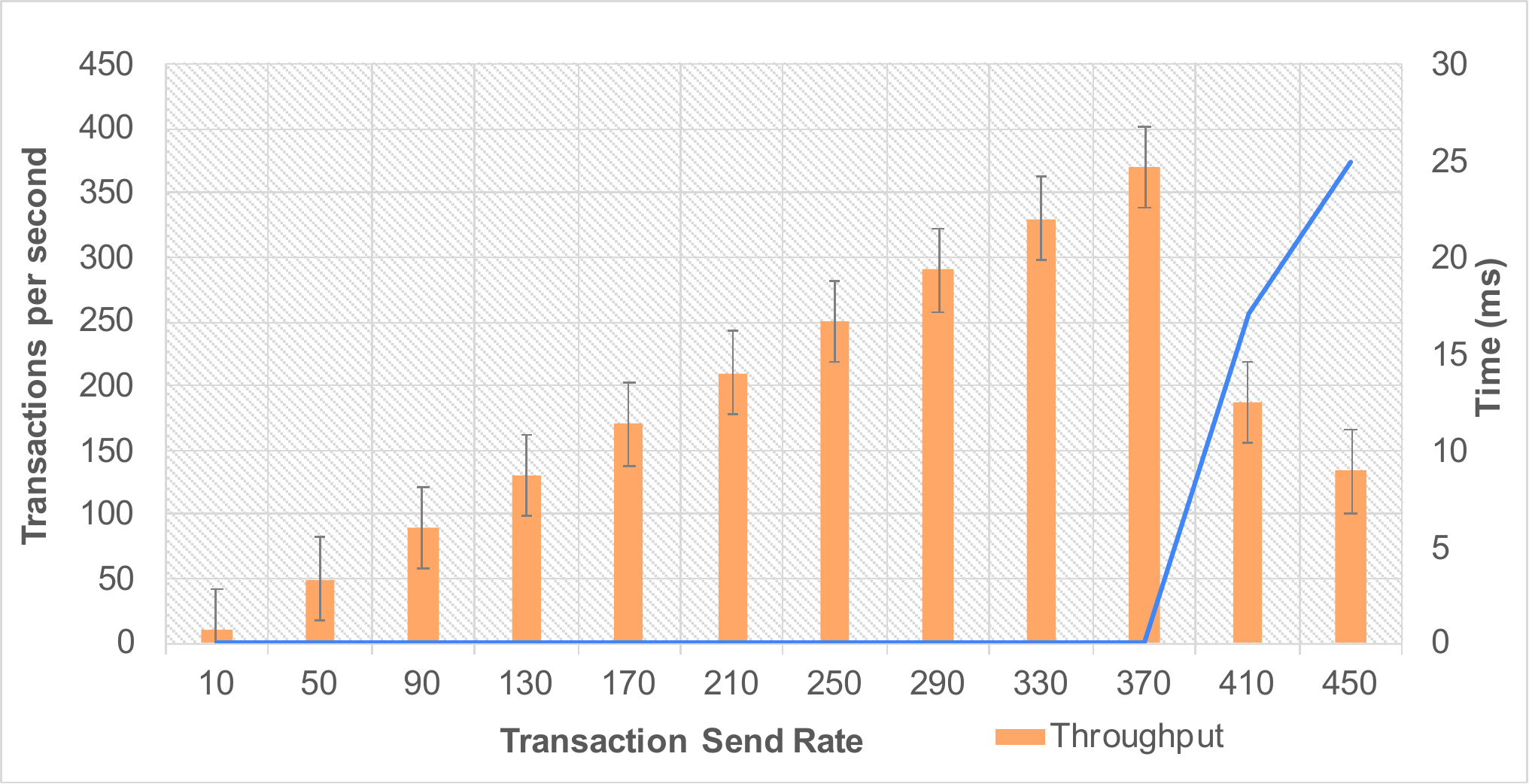}}     &
\resizebox{0.42\textwidth}{!}
{\includegraphics{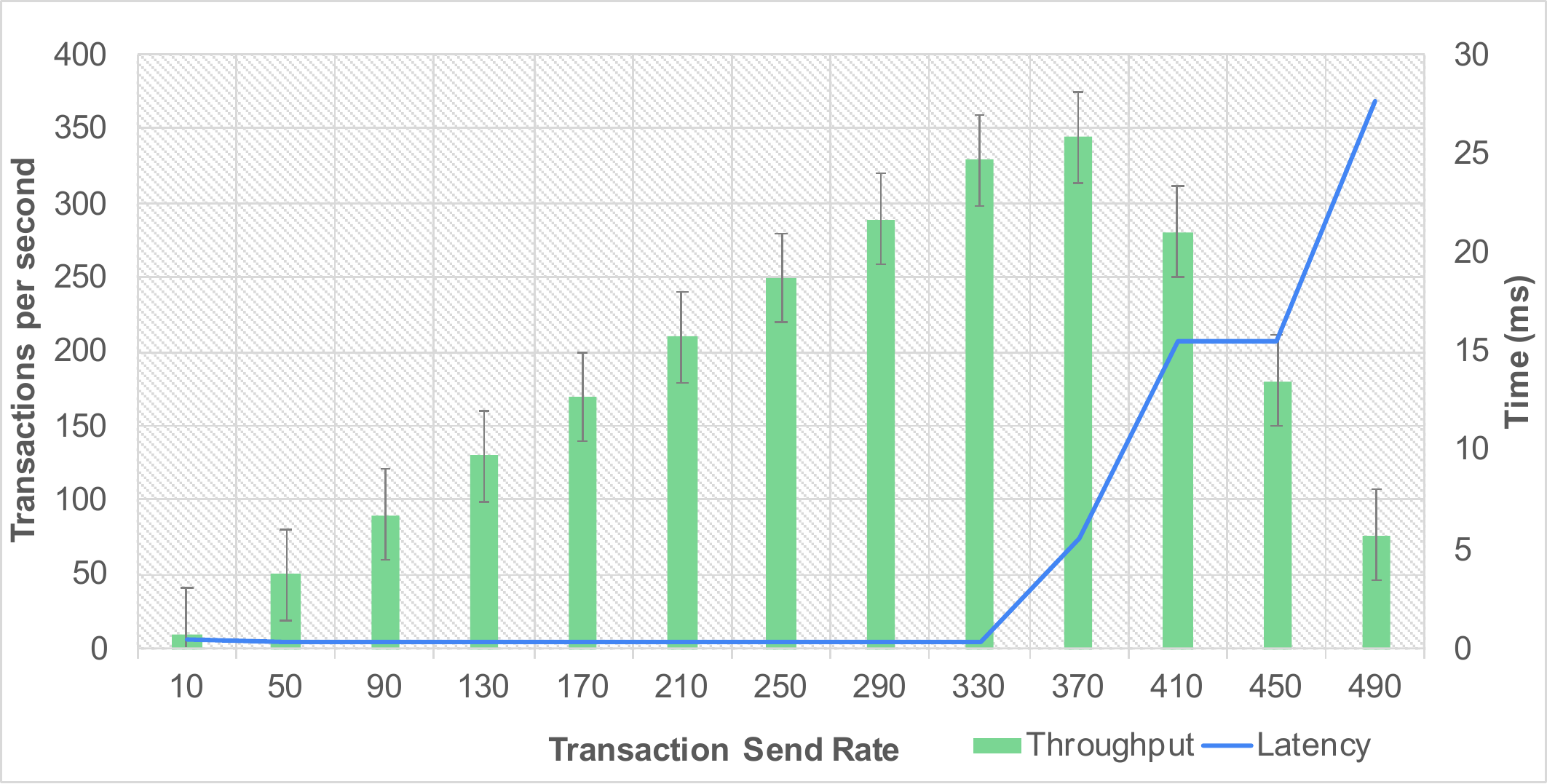}}  \\

{(a) \small{Registering a commodity $TX_{cr}$}} & 
 {(b) \small{Trading a commodity $TX_{tr}$}} \\

\resizebox{0.42\textwidth}{!}
{\includegraphics{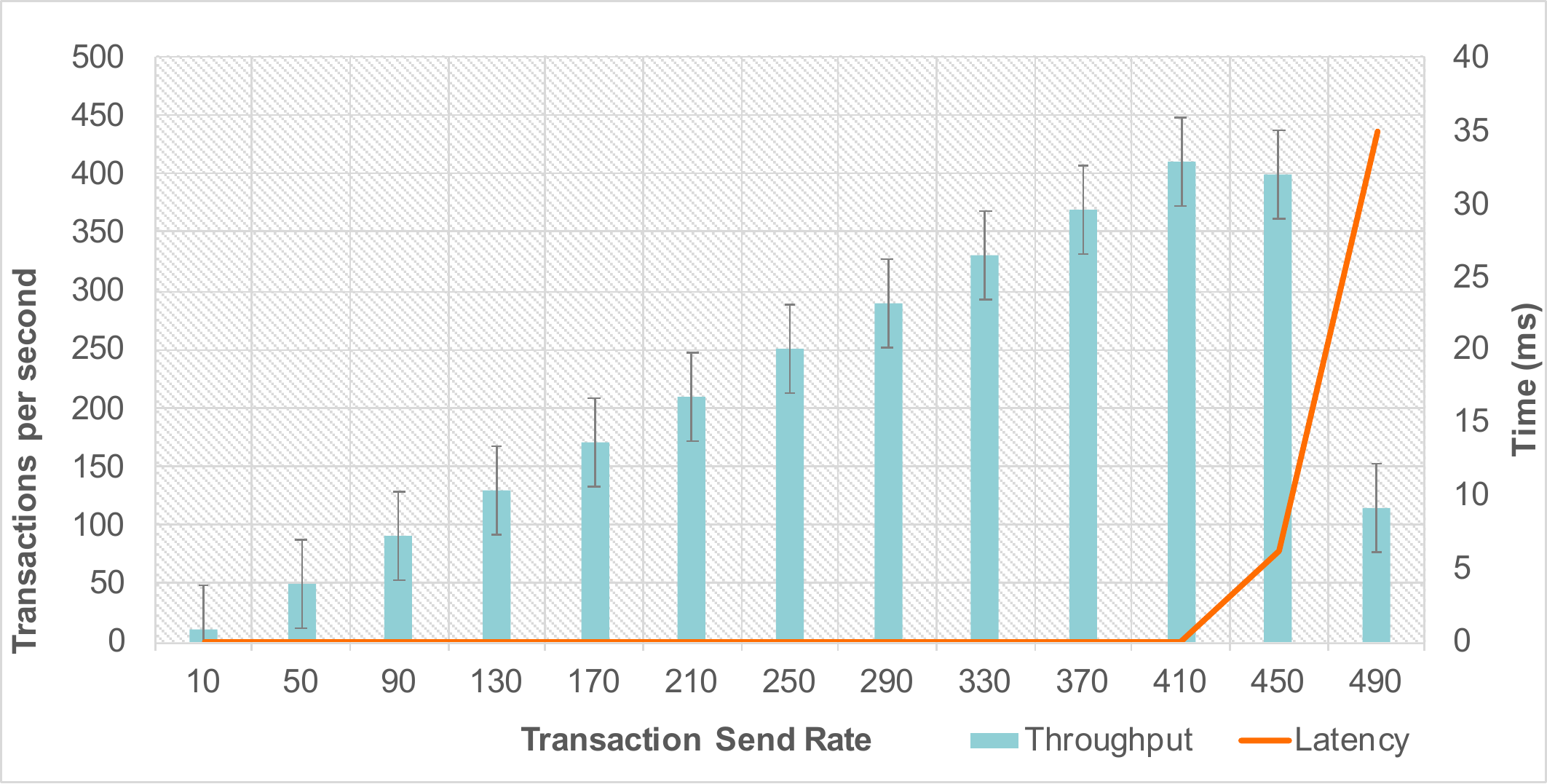}}  &
\resizebox{0.42\textwidth}{!}
{\includegraphics{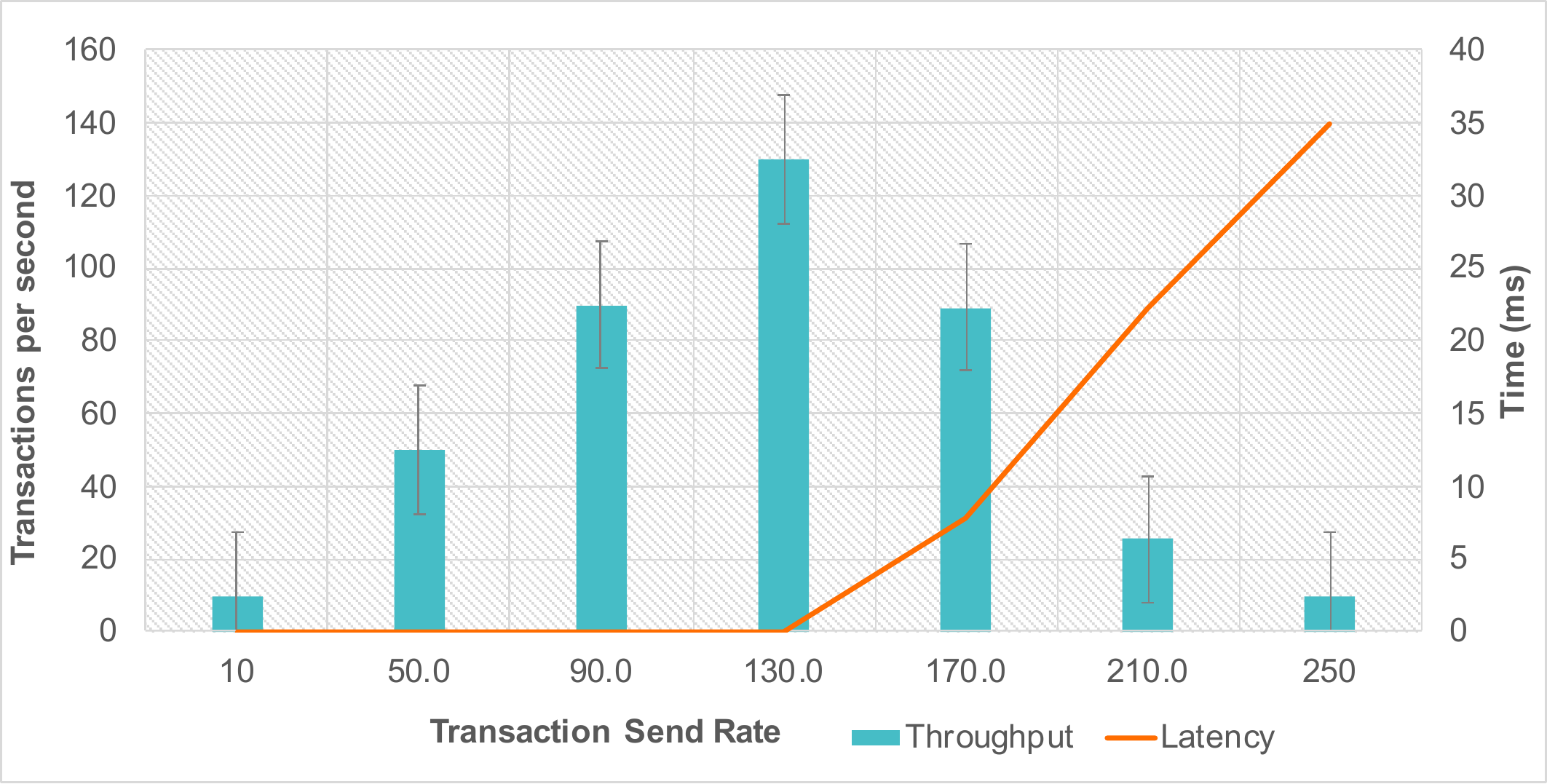}}   \\

{\small{(c) Querying TXns based on token}} &
{\small{(d) Querying and Returning filtered results}} \\

\end{tabular}
\end{center}

\caption{Throughput and Latency of TradeChain}
\label{fig:caliperres}
\end{figure*}

For the latency and throughput computations on TML, the transaction send rate is varied from 10 to 500 transactions per second (tps). The cumulative results are shown in Figure \ref{fig:caliperres}. Recall from Section \ref{sec:TML}, each commodity is registered on TML using $TX_{cr}$ and traded using $TX_{tr}$. Figure \ref{fig:caliperres}-a and Figure \ref{fig:caliperres}-b shows the throughput and latency for $TX_{cr}$ and $TX_{tr}$ respectively. The throughput increases linearly till the transaction send rate of 370 tps for $TX_{cr}$ and 410 tps for $TX_{tr}$ and declines after. For both transactions, the latency increases after the transaction send rate of 330 tps. $TX_{tr}$ has a higher throughput than $TX_{cr}$ because it only changes the ownership status of an existing commodity on TML whereas in $TX_{cr}$, a new commodity is generated which adds to the delay.\\


In the next set of evaluations, we evaluate the transactions related to access tokens. 


Figure \ref{fig:caliperres}-c  shows the computations for decoding the access token and returning all the transactions related to query parameters of the token. When the transaction results are further filtered to remove $DID_p$s and parameters that are not part of the access token, the maximum throughput of TradeChain is 130 tps as shown in Figure ~\ref{fig:caliperres}-d. The latency increases and throughput decreases after this point. This is because in Figure~\ref{fig:caliperres}-c, the transactions are just being retrieved from the ledger. However, in Figure \ref{fig:caliperres}-d, a significant increase in latency is observed due to the need to filter the results according to access token parameters. Given that the maximum throughput is only 130 tps for filtered queries, querying using the access token has the lowest throughput among all the system components. We believe that these performance metrics should meet the needs of most real-world scenarios because the number of requested transactions for an audit is usually not time-sensitive i.e. they do not incur a high transaction rate and are expected to be well below 130 queries per second. 

\subsection{Security and Privacy Analysis} 
In this section, we briefly explain our threat model and analyse the major security threats and TradeChain's resilience towards them.\\
\textbf{Threat Model:}
Given that the traders could use multiple DIDs to obscure their trade related information, some dishonest traders may use this as an opportunity to hide their identities while performing malicious activities. The information requesters may also attempt to query additional query parameters from the ledger apart from those mentioned in the access token. In our threat model, we consider how dishonest traders and information requesters could execute some malicious actions to undermine various functions of the TradeChain framework. The considered attacks include:
\subsubsection{Creating Multiple DIDs (DOS)}
A malicious trader may create multiple $DID_v$s without using them for transactions. Similarly, a number of $DID_p$s can be generated in excess but requires a trader to collude with another trader ( recall from section~\ref{sec:onboard} that $DID_p$s are generated in pairs). A large number of transactions registering $DID_v$s on IDML may increase the network load. \par 
To counter this, there are two checks: in IDML, every trader has a threshold for $DID_p$ generation in a given period of time. Even if the traders collude to generate $DID_p$s, this can be detected. In TML, a trader can create a new $TX_{cr}$ using a new $DID_v$ only if the previously created $DID_v$ was used in a transaction. In this way, the traders can be restricted for DID usage in both ledgers.

\subsubsection{ $DID$ and Wallet Deletion}
A trader may wish to delete specific $DID_p$s to delete tracking of his trades in TML. If a trader is allowed to delete his wallet completely, then the associated transactions in IDML and TML cannot be traced at the time of audit. \par
The digital wallets of the supply chain participants are a part of Hyperledger Indy SDK. These wallets are append-only. Thus, a trader cannot delete specific DIDs. If a trader wishes to delete a wallet, his wallet credentials can be revoked by the Steward, and each DID can be tagged as `deleted' without the actual removal of the DID. This allows the information in TML to be still collated. Alternately, we may allow transactions to be removed from TML while maintaining verifiable copies off-chain \cite{dorri2019mof}. The transactions could be marked as ``aged" after a product's estimated life-cycle or expiry. The $DID_p$s for aged transactions can then be tagged as deleted. 

\subsubsection{Linking Trades using CID}
The $CID$ field is the commodity identifier which is used in $TX_{cr}$ (see Eq. 2) and $TX_{tr}$ (see Eq. 3). 
For traceability purposes, $CID$ is kept in plaintext. Recall the identity of the owner field in $TX_{cr}$ is $DID_v$. When the commodity is traded using $TX_{tr}$, the owner field is updated using $DID_p$ of the seller. This links the $DID_{p}$ of the seller in the first $TX_{tr}$ of the commodity to the $DID_v$ in $TX_{cr}$. Since the traders can create new DIDs for each commodity, the commodities belonging to same seller cannot be linked. 
However, the $CID$ field can optionally be encrypted, and a smart fingerprint can be created such as proposed in \cite{laava}, without losing instant traceability and provenance. 


\subsubsection{Access Token Modification}
An information requester may request a token for certain parameters and after it has been issued, he may try to modify the token by adding more query parameters.\par
Signed hash of token and $Param_i$ by the trader issuing the token, ensures that the original query parameters are not changed. Hence, it is impossible that a token is modified without the trader's consent. 
\section{Conclusion}
In this work, we have presented TradeChain, an architecture for decoupling identities and trade activities on blockchain enabled supply chains. The proposed architecture utilises the decentralised identifiers maintained on a separate ledger, and allows traders to log trade events on a permissioned ledger by providing ZKPs on these identifiers. In addition, we have proposed a token based query mechanism to collate information from both ledgers, and retrieve trade history of an individual trader. These access tokens are encrypted using CP-ABE by traders themselves, allowing the information from the ledger to be only retrieved with the trader's consent. We demonstrated the feasibility of TradeChain by implementing a proof of concept implementation on Hyperledger Fabric and Indy. The performance metrics of Tradechain are presented as time overheads, throughput and latency. We have also analysed privacy and security aspects of TradeChain which shows its resilience to malicious intent of traders and information requesters. 
\ifCLASSOPTIONcaptionsoff
  \newpage
\fi

\bibliographystyle{IEEEtran}
\bibliography{mybib}

\end{document}